\documentstyle[11pt,psfig]{article}

\newcommand{\be}{\begin{equation}}
\newcommand{\ee}{\end{equation}}
\newcommand{\beq}{\begin{eqnarray}}
\newcommand{\eeq}{\end{eqnarray}}
\newcommand{\bed}{\begin{displaymath}}
\newcommand{\eed}{\end{displaymath}}
\begin{document}
\baselineskip 1.2\baselineskip
\thispagestyle{empty}

\begin{center}

\vspace{2.0in}

{\Large\bf{A Cosmological Model of Holographic Brane Gravity}}

\vspace{1.0 in}

Piret Kuusk$^{1}$ and Margus Saal$^{2}$\\
Institute of Physics, University of Tartu,
Riia 142, Tartu 51014, Estonia\\

\end{center}

\vspace{0.5in}

\begin{abstract}

A cosmological scenario  
with two branes (A and B) moving in a 5-dimensional bulk
is considered. As in the case  
of ecpyrotic and born-again braneworld models it is possible that
the branes collide. 
The energy-momentum tensor  is taken to describe
a perfect barotropic fluid on the A-brane and  a phenomenological 
time-dependent "cosmological constant" on the B-brane.  
The A-brane is identified with our Universe and its cosmological 
evolution in the approximation of a homogeneous and isotropic brane 
is analysed. 
The dynamics of the radion (a scalar field on the brane) contains information 
about the proper distance between the branes.   
It is demonstrated that the deSitter type solutions are obtained 
for late time evolution 
of the braneworld and accelerative behaviour is anticipated  at the present 
time.

\end{abstract}

\vspace{0.3 in}

%\pacs{98.80 Dr}

\vspace{0.5in}

$^1$ Electronic address: piret@fi.tartu.ee

$^2$ Electronic address: margus@hexagon.fi.tartu.ee

%%%%%%%%%%%%%%%%%%%%%%%%%%%%%%%%%%%%%%%%%%%%%%%%%%%%%%%%%%%%%%%%%%%%%%%%

\newpage

\section {Introduction} \label{uks}

To solve the hierarchy problem Randall and Sundrum \cite{rs1}, \cite{rs2}
 proposed a scenario 
 where our spacetime is a 4-hypersurface (a world volume of a 3-brane)
 in a 5-dimensional bulk 
spacetime with $Z_2$ symmetry along the extra dimension. 
They made two different proposals. The first RS1 \cite{rs1} has two branes
of opposite tensions in an anti-deSitter (AdS) 
background spacetime and the second 
RS2 \cite{rs2} has a single positive tension brane.   
The cosmology was analysed on both occasions \cite{bdl}, 
for a review see \cite{langlois}.

Kanno and Soda derived  low energy effective actions for 
one and two brane systems \cite{ks2}, \cite{ks1}. 
The latter one looks like a scalar-tensor theory of 
gravity on braneworlds,
which is characterized by a distinct coupling function.  
The proper distance between the branes in 
a 5-dimensional spacetime is parametrized by radion 
$\phi$
and the radion field 
appears as a scalar field on our brane, which 
allows us to write 
effective 4-dimensional field equations in a scalar-tensor form with an 
additional 
term describing the influence of the second brane. 
We assume that this scalar-tensor theory of gravity describes the dynamics 
of our 3-brane Universe and is therefore applicable to analyse 
the cosmological evolution on the brane. 
If we know the dynamics of the 3-brane, it is possible, 
through a holographic 
conjecture, say something also about 5-dimensional bulk as 
pointed out by Kanno and Soda \cite{ks2}. 
 
We keep in mind the general idea of the ecpyrotic 
(and possibly cyclic) model of the brane 
Universe introduced by Khoury, Ovrut, Steinhardt, and Turok \cite{ekpyrotic} 
and analyse the 4-dimensional field equations, derived from the 5-dimensional 
theory, in this context.
In the ecpyrotic  model, the Universe is initially contracting 
towards a big 
crunch and then makes a transition through a singularity
to the post-Big Bang Universe. The Big Bang (the initial singularity) 
is treated as a 
collision of branes in a 5-dimensional AdS spacetime. 
However, the problem of singularity remains \cite{ks}, 
since it is very difficult to get rid of the
singularity and to mix/incorporate the ecpyrotic
and the pre-Big Bang \cite{pbb}, \cite{homepage} models.

Recently Kanno, Sasaki and Soda \cite{kss} have proposed
a new type of braneworld cosmology, so-called born-again braneworld (BAB).
In the BAB scenario the branes are empty in the sense that
they carry no nontrivial energy-momentum tensor. It is supposed that
the signs of tensions
of branes are changed after the collision. 
The model implies a cosmology which resembles  the pre-big bang 
scenario \cite{pbb}, \cite{homepage} in some respects.

In this paper we accept the general setting of the BAB scenario
and investigate the field equations derived from the quasi-scalar-tensor 
theory on the branes.   
We assume that  energy-momentum tensors on the branes are nontrivial
and describe 
a perfect barotropic fluid on the A-brane and  a phenomenological 
time-dependent "cosmological constant" on the B-brane.  
We analyse how the 4-dimensional dynamics on the A-brane is 
influenced by the motion of the B-brane and by the B-brane matter.
A special attention is paid to the dynamics of the radion. 
A possibility of nontrivial contribution of the bulk geometry, 
so-called dark radiation on the A-brane,
is investigated. 

The paper is organized as follows. 
In the next section, the general field equations derived by Kanno and 
Soda \cite{ks1} together with the gradient expansion formalism 
in the leading order are presented.
In the third section,  the field equations are solved in the case 
of vanishing dark radiation   
and solutions are analysed in the context of ekpyrotic and BAB models.
In the fourth section, the influence of the dark radiation on the
cosmological evolution on the A-brane is considered.
The fifth section is a summary.

\section {Field equations} \label{kaks}

Our starting point is a RS1 type model with 
two 3-branes at the orbifold fixed points.
We put a positive tension brane (A-brane) at $y=0$ and a negative tension 
brane (B-brane) at $y=l$ and describe the 5-dimensional spacetime
with the metric \cite{ks1}
\begin{equation}
ds^2 = e^{2\phi (y,x^{\mu})} dy^2 
+ g_{\mu\nu} (y,x^{\mu} ) dx^{\mu} dx^{\nu}  \ , 
\end{equation}
where radion field $\phi(y,x^{\mu})$ measures the proper 
distance between the branes 
$d(x) = \int_0^l e^{\phi (y,x)} dy$. As stated by Kanno and Soda \cite{ks1},
it is possible to choose the coordinate system to be such that
$\phi (y,x) = \phi (x)$; then $d(x) = l e^{\phi (x)}$ which implies 
$d(x) \geq 0$. 

The general 5-dimensional action can be taken as follows
\beq \label{5moju}
 S &=& {1\over 2\kappa^2}\int d^5 x \sqrt{-g} \left( 
 {\cal R} + {12\over \ell^2} \right)  
 -\sum_{i=A,B} \sigma_i \int d^4 x \sqrt{-g^{i{\rm brane}}}\nonumber\\
 &+& \sum_{i=A,B} \int d^4 x \sqrt{-g^{i{\rm brane}}} 
 {\cal L}_{\rm matter}^i \ ,
\eeq
where ${\cal R}$ is the scalar curvature,  $g^{i{\rm brane}}_{\mu\nu}$ are 
induced metrics on branes, $\kappa^2  $  is the five-dimensional 
gravitational constant, $\sigma_A=6/(\kappa^2\ell)$, 
$\sigma_B=-6/(\kappa^2\ell)$ 
are the tensions of branes and $\ell$ is the curvature radius of 
5-dimensional AdS bulk spacetime\footnote{Randall and Sundrum \cite{rs1}
used a notation $2\Lambda = \frac{12}{\ell{^2}}$, where $\Lambda$ 
is the 5-dimensional 
cosmological constant and $\sigma_{A} = V_{vis}$ , $\sigma_{B} = V_{hid}$.}.  
Using this setup and the low energy expansion scheme in the sense that 
energy density of the matter on the brane is much smaller than the brane 
tension $\rho_i /|\sigma_i| \ll 1$, Kanno and Soda derived \cite{ks1} 
the 4-dimensional effective equations for the A-brane
\beq \label{einstein}
R^\mu_{\ \nu} (h) &=&{\kappa^2 \over \ell \Psi } \left(T^{A\mu}_{\quad\ \nu}
-\frac{1}{2}\delta^\mu_\nu T^{A} \right) 
+{\kappa^2 (1-\Psi ) \over \ell\Psi } \left(T^{B\mu}_{\quad\ \nu}
-\frac{1}{2}\delta^\mu_\nu T^{B} \right) \nonumber\\
&+&{ \frac{1}{2 \Psi} } \delta^\mu_\nu \Box \Psi + \frac{1}{\Psi} 
\Psi^{|\mu}_{\ |\nu} 
+{\omega(\Psi ) \over \Psi^2}  \Psi^{|\mu}  \Psi_{|\nu}  \ ,
\eeq

\be \label{dilaton}
\Box \Psi = {\kappa^2 \over \ell} {T^A + T^B \over 2\omega +3}
  		-{1 \over 2\omega +3}{d\omega \over d\Psi} \Psi^{|\mu} 
  		\Psi_{|\mu} \ .
\ee
Here $T^{A\mu}_{\quad\ \nu}$ and $T^{B\mu}_{\quad\ \nu}$ are the 
energy-momentum tensors of the A-brane and the B-brane, respectively,  
$h_{\mu\nu}(x) = g_{\mu\nu}(y=0,x)$ 
is the induced metric on the A-brane and $|$ is the derivative with 
respect to the A-brane metric $h_{\mu\nu}$.
The scalar field $\Psi$ is determined by the radion field  $\phi$ as 
follows  
\begin{equation}  \label{psii}
\Psi = 1- \exp(-2e^\phi) 
\end{equation}
which implies $\Psi \in [ 0, 1 ]$.
The coupling function $\omega(\Psi)$ reads 
\begin{equation} 
\omega (\Psi ) = {3\over 2} {\Psi \over 1-\Psi }  
\end{equation}
and must be  substituted into  field equations (\ref{einstein}), 
(\ref{dilaton}).
As distinct from the situation in a general scalar-tensor type theory we
don't need any additional {\it{ad hoc}} hypothesis about the 
form of the coupling function here.

In the limit $\Psi \rightarrow 1$ we get the familiar general relativity and
this corresponds to a situation of a large distance between the branes 
\begin{equation} 
d(x) =  l e^\phi =- \frac{l}{2} \ln (1-\Psi) \rightarrow  \infty,
\qquad  \phi \rightarrow + \infty \ .
\end{equation}
The other limit $\Psi \rightarrow 0$ corresponds to a situation where 
the branes collide: $d \rightarrow 0, \quad \phi \rightarrow - \infty$.

The conservation of energy-momentum tensor with respect to  induced 
metrics gives  additional constraints \cite{ks1}
\be \label{js1}
   	T^{A\mu}_{\quad\ \nu |\mu } =0 \ , \quad
   	T^{B\mu}_{\quad\ \nu |\mu } = {\Psi_{|\mu} \over 1-\Psi }
        T^{B\mu}_{\quad\ \nu} 
        -{1\over 2}{\Psi_{|\nu} \over 1-\Psi}T^{B} \ .
\ee

In what follows we analyse  field equations (\ref{einstein}), 
(\ref{dilaton}) 
in the case when spatial gradients and local anisotropy are absent.
For the A-brane we assume a perfect fluid matter
\be \label{pfluid}
T^{A}_{\quad \mu\nu} = (\rho + p) u_{\mu} u_{\nu} + p g_{\mu\nu}\ , \quad 
p = (\Gamma - 1) \rho \ , \quad 0 \leq \Gamma \leq 2 \ 
\ee
and for the B-brane we take the energy-momentum tensor in the simplest 
non-trivial form{\footnote{This 
corresponds to $L^{B} = -\delta \sigma^{B}$ in the case of 
the BAB scenario \cite{kss}.}}
\be
T^{B\mu}_{\quad \nu} = \lambda^B (t) \delta^{\mu}_{\nu} \ . 
\ee

Let us  introduce a synchronous gauge on the A-brane
\be
 ds^{2} = -dt^{2} + h_{ij}(t,x^{k}) dx^{i} dx^{j}
\ee  
and assume that a solution for the A-brane can be taken in a 
quasi-isotropic form
\beq \label{qi}
 h_{ij}(t, x^{k}) = a^{2}(t) \ f_{ij}(x^{k}) \ . 
\eeq
Here $f_{ij}(x^k)$ is a time independent seed metric and $a(t)$  
is the scale factor of an isotropic and homogeneous A-brane Universe.

Constraints (\ref{js1}) imply the usual form of the conservation 
law for perfect fluid matter
\be \label{pidevus}
\dot \rho + 3 H \Gamma \rho = 0
\qquad \Rightarrow \qquad \rho = \rho_{0} a^{-3\Gamma} 
\ee
and a simple equation for $\lambda^B$
\be \label{l1}
\frac{\dot \lambda^B}{\lambda^B}  = - \frac {\dot \Psi}{ ( 1 - \Psi)} \ .
\ee
Here $H = {\dot a}/ a$ is the Hubble parameter on the A-brane 
and dot means the derivative with respect to time $t$ in the 
synchronous gauge. 

The solution of equation (\ref{l1}) reads
\be
\lambda^{B} = \lambda_{0}^{B}(1 - \Psi) = \lambda_{0}^{B}\exp(-2e^\phi) \ .
\ee
As we can see the evolution of the  B-brane ``cosmological 
constant" with respect to the A-brane synchronous time $t$ is in fact 
parametrized by the proper distance (radion) between the branes.
If $\phi$ is large (the distance between the branes is large) 
 the ``cosmological constant" of the B-brane almost vanishes 
($\lambda^{B} \rightarrow 0$) 
and {\it vice versa}, if the distance between the branes is small, then 
the ``cosmological constant" on the B-brane has a nonvanishing value.

Now, using  Ansatz (\ref{qi}) and ignoring  spatial derivatives 
we can write the 4-dimensional  field equations 
(\ref{einstein}) and (\ref{dilaton}) as follows 

\beq \label{v1}
\ddot \Psi + 3 H \dot \Psi + \frac{1}{2} \frac{\dot \Psi^2}{(1-\Psi)} =
\frac{ \kappa^2}{3 \ell}(1-\Psi) (4 - 3\Gamma ) \rho -  
\frac{4 \kappa^2}{3 \ell} (1 - \Psi)^2 
\lambda_{0}^{B} \ ,\\ \nonumber \\ 
\label{v2}
\dot H + 2 H^2 = \frac{\kappa^2}{6 \ell}(4 - 3\Gamma ) \rho \ ,\\ \nonumber \\ 
\label{v3}
H^2 +  H \frac{\dot \Psi}{\Psi} - 
\frac{1}{4} \frac{\dot \Psi^2}{\Psi (1 - \Psi)} = 
\frac{\kappa^2}{ 3 \ell} \frac{\rho}{\Psi}  - \frac{\kappa^2}{ 3 \ell}  
\frac{(1-\Psi)^2}{\Psi} \lambda_{0}^{B} \ .  
\eeq
Equations (\ref{v1}) and (\ref{v2}) are  dynamical equations 
for $\Psi$ and 
$H$ respectively and (\ref{v3}) is a generalization of the 
Friedmann equation. Here we treat it as an additional constraint. 
As we can see,   equation for $H$ (\ref{v2}) 
does not contain any additional 
terms describing the influence of the B-brane and the scalar field. 
It is exactly the same equation we have in 
the Einstein general relativity without 
any additional scalar field and the B-brane ``cosmological constant''.   
Equation (\ref{v1}), on the contrary, contains an additional term 
on the l.h.s. compared 
with the scalar-tensor type theories which introduces a strong 
non-linearity into the equation. 
In what follows, we present and analyse some special solutions of these
equations.

\section{Solutions with vanishing dark radiation}

It seems very convenient to assume  $\Psi = 1 - e^{-2 m t}$, 
where $m$ is a constant whose value should be found from the equations.
This kind of solution satisfies an important condition, namely, 
at late times, 
when the distance between the branes is large, we must  effectively 
get the usual 
general relativity on the brane (i.e. $\Psi \approx 1$). But
this choice leads to unacceptable conditions for $\Gamma$ and  
doesn't satisfy 
the constraint equation (\ref{v3}).

In the absence of the B-brane (i.e. $ \lambda_{0}^{B}=0$),  
equations (\ref{v1})--(\ref{v3}) coincide with  equations 
of a scalar-tensor theory treated by Serena et al \cite{sa1}. 
Unfortunately we cannot use their procedure for 
finding a general solution of our equations, 
because it is not possible to eliminate simultaneously 
energy density on the A-brane and  
''cosmological constant'' on the B-brane 
from the equations.
However, we have succeeded in solving the equations in some nontrivial 
special cases.

\subsection{$\Psi = const$}

If we assume that the proper distance between the branes doesn't change 
and $\Psi =1$, then eq. (\ref{v1}) is trivially satisfied and eqs. 
(\ref{v2}), (\ref{v3}) reduce to the familiar FRW equations. 

If $\Psi = {\tilde \Psi}_0 = const \not= 1$ we get from
equation (\ref{v1}) a constraint between the sources
\be \label{rl}
\rho = \frac{4 (1 - {\tilde \Psi}_0)}{(4-3\Gamma)} \lambda_{0}^{B} = const \ .
\ee
As we can see from the conservation law (\ref{pidevus}), 
in the case of a nonstatic ($H \not= 0$) nonempty ($\rho \not= 0$)
Universe the constant energy density 
implies  $\Gamma = 0$ and  
\be
\rho = \tilde{\rho}_{0} = (1 -{\tilde \Psi}_0 )\lambda_{0}^{B} \ .
\ee 
For the radion field we get  
\be
{\tilde \Psi}_0= 1 -\frac{\tilde \rho_0}{\lambda_{0}^{B}}, \qquad
e^{\phi} = -\frac{1}{2} \ln (1 - {\tilde \Psi}_0) \ .
\ee 
This situation corresponds to the case of a
"phenomenological cosmological constant" ($p = - \rho$) on the A-brane. 
The constraint equation (\ref{v3}) determines the Hubble parameter as
\beq \label{H^2}
H^2 = k \tilde \rho_{0} =  k \lambda_{0}^{B}(1- {\tilde \Psi}_0) \ , 
\qquad k = \frac{\kappa^2}{3 \ell}
\eeq
which leads
to  the deSitter type inflation on the A-brane
\be \label{dS}
a = a_{0} e^{\sqrt{ k \tilde \rho_{0}} t} 
\ee
during the time when the proper distance between the branes is constant 
in a static bulk.

Now we give a different point of view to this solution.
We can write  equations (\ref{v1})--(\ref{v3}) in a form  
of  a dynamical system as follows
\beq \label{Htapp}
\dot H & = & -2 H^2 + \frac{1}{2} k \rho (4 - 3\Gamma)  \ , \\
\label{ptapp}
\dot \Psi & = & 2 H (1 - \Psi) \pm 2 \sqrt{1-\Psi} 
          \sqrt{H^2 - k\rho + k \lambda_{0}^{B} (1 - \Psi)^2 } \ . 
\eeq
Conditions for an equilibrium point of the dynamical system are 
$\dot H =0, \quad \dot\Psi = 0$. If we take $\Gamma = 0$ we get 
the same relation (\ref{H^2}) for $H^2$ 
and the same 
constraint (\ref{rl}) for $\rho$,  $\lambda_{0}^{B}$  as before, 
which indicates that the exponential 
expansion and the constant dilaton is an equilibrium point of the system
(\ref{Htapp})--(\ref{ptapp}). 
If we keep $\Gamma$ general at the beginning, then
the first equation (\ref{Htapp}) implies $\rho= const$ 
at the equilibrium point   and the 
 conservation law (\ref{pidevus}) implies once again $\Gamma =0$.

\subsection{$\Gamma = 0$,  $\Psi \not= const$}

The conservation law (\ref{pidevus}) with  $\Gamma = 0$ implies  
$\rho = \rho_0= const$ and (\ref{v2}) acquires a form of 
the familiar FRW equation
\be \label{FRW}
{\dot H} + 2 H^2 = 2k \rho_0 \ .
\ee
It is trivially satisfied by the
constant Hubble parameter $H^2 = k \rho_{0}$ (however, this is not 
its general solution, see Sect. 4). 
For the scale factor we get the exponential evolution as before
(\ref{dS}).
Constraint equation (\ref{v3}) for $\Phi = 1 - \Psi$ now  reads 
\be
\dot \Phi^2 + 4 \sqrt{k \rho_{0}} \Phi \dot \Phi + 
4 k \rho_{0} \Phi^2  - 4 k \Phi^3  \lambda_{0}^{B} = 0 \ . 
\ee 
Solving it as an algebraic equation for ${\dot \Phi}$
and then as a 
differential equation we get for $\Psi$
\be \label{g0psi} 
\Psi = 1 - \frac{ \rho_{0}}{\left(A \sqrt{ \rho_{0}} e^{\sqrt{k \rho_{0}} t} 
\pm \sqrt{ \lambda_{0}^{B}}\right)^2} \ . 
\ee
Here $A$ is an integration constant and if we choose  
$A = \sqrt{ \lambda_{0}^{B}/ \rho_{0}}$, then
\be \label{p1}
\Psi = 1 - \frac{ \rho_{0}}{ \lambda_{0}^{B}} 
\frac{1}{\left(  e^{\sqrt{k \rho_{0}} t} \pm 1 \right)^2} \ .
\ee
If we choose $A = 0$ we get the same solution $\Psi= const$ 
as in the previous subsection 3.1.

Solution (\ref{g0psi}) determines the proper distance between branes
\be
d(t) = l e^{\phi (t)} = l \ln \bigl(A e^{\sqrt{k \rho_{0}} t} \pm  
\sqrt{\frac{ \lambda_{0}^{B}}{ \rho_{0}}}
 \bigr)  \ .
\ee
Let us analyse it more in detail. We have
\beq
 t \rightarrow +\infty \qquad &d& \rightarrow + \infty \ , \\ 
t=t^\pm_c= \frac{1}{\sqrt{k\rho_0}} \ln 
\frac{\sqrt{\rho_0} \mp \sqrt{\lambda^B_0}}{ A \sqrt{\rho_0}}
\qquad    &d&=0 \ .
\eeq
However, solution (\ref{g0psi}) has no singularity at the
moment when $d=0$ (collision of branes) and can be continued in
a region where $d(t) < 0$. This can be interpreted as a change in  
the sequence of branes along the $y$-axis: $d(t) > 0$ for a sequence  (A,B) 
and $d(t) < 0$ for a reversed sequence (B,A). Note that  the
BAB scenario includes  a reversed
sequence rather naturally. 
In this case we have $d=- \frac{l}{2} \ln (1-\Psi) \leq 0$, which 
corresponds to an extension of the domain of
values of $\Psi$  by  $\Psi \in (-\infty, 0 )$.

Let us investigate the evolution of the proper distance at $d(t) < 0$
   separately for both signs.

In the case of "--",
if $t^-=0$ then  $d \rightarrow -\infty$. We see that 
the  proper distance between branes covers the whole
real axis $d \in (-\infty, +\infty)$
 during $t \in [0, + \infty)$,  the solution is not determined
at negative times and the scale factor $a(t)$ never vanishes. 

 In the case of "+" we have
\be
t^+ \rightarrow -\infty  \qquad  
d = l \ln \sqrt{\frac{\lambda_{0}^{B}}{\rho_{0}}} <0 \ . 
\ee
We see that at infinitely remote past the Universe contains two branes 
at a finite distance and the exponential scale factor (\ref{dS}) 
of the A-brane  tends to zero (our Universe has a  singularity at
$t^+ \rightarrow -\infty$).  
At the moment of the collision of the branes  $a(t)$ is perfectly regular.
The situation is depicted on Fig. 1, where    
the constants are chosen as follows: 
$A = \frac{\sqrt{ \lambda_{0}^{B}}}{\sqrt{  \rho_{0}}} = \frac{1}{4}$ , 
$\sqrt{ k \rho_{0}} = 0.4$ and ${\sqrt{ k \lambda_{0}^{B}}} = 0.1$ \ .

\begin{figure}[h]
\centering
\vskip 1cm
\hskip 1.5cm 
\psfig{file=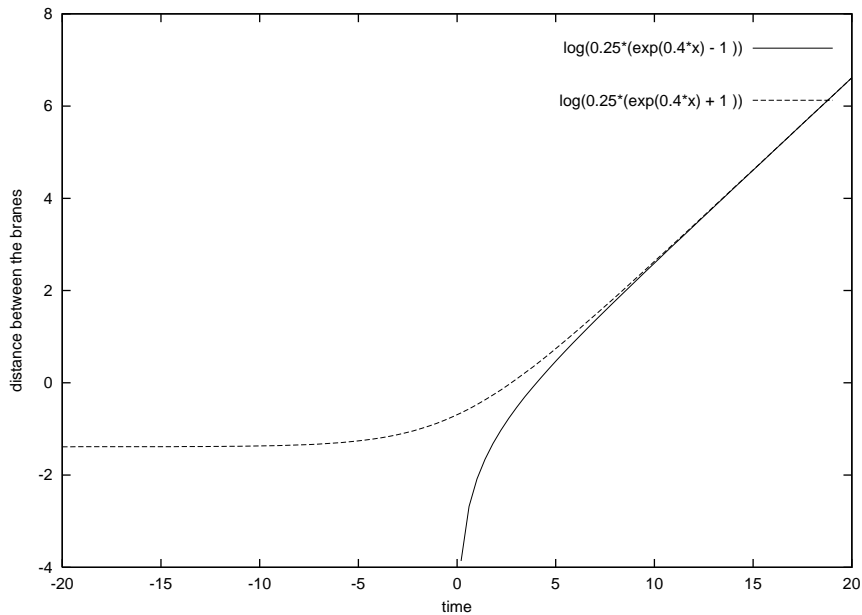}
%\vskip 1cm
\caption{The proper distance between the branes in the case of $\Gamma =0$}
\label{figure1}
\end{figure}

\subsection{ $\Gamma \not= 0$}

Let us assume, that the evolution of the scale factor on the A-brane 
is described by a power function $a = a_{0} t^m$ as in a typical 
case of general relativity \cite{kk}. From 
equation (\ref{v2}) and the conservation law (\ref{pidevus}) 
we get an expression for the power index 
$m = \frac{2}{3\Gamma} , \ \Gamma \not= 0$ and a 
constraint for  initial values
$k \rho_{0} = \frac{4}{9 \Gamma^2} (a_{0})^{3\Gamma}$.
As a result the solution for the scale factor coincides with
the familiar FRW solution
\be \label{ah} 
a(t) = a_0 t^{\frac{2}{3 \Gamma}} \ , \qquad H= \frac{2}{3\Gamma t} \ .
\ee
Now the  energy density on the A-brane reads  
$\rho = \rho_{0} (a_{0})^{-3\Gamma} t^{-2}$   
and we get 
\be \label{h2}
H^2 (t) = k \rho (t) \ .
\ee
This is just the familiar Friedmann equation: 
the assumption of the power law evolution on the A-brane 
reduces the generality of the solution which can be interpreted as 
ignoring the contribution of the dark energy;  
that's why we obtained 
the same constraint (\ref{h2}) as in general relativity.  
Note that  solution (\ref{ah})
is not singularity free and accepting it means to accept 
and not to solve the problem of singularity.

Taking into account expressions (\ref{ah}), (\ref{h2}) 
we get from constraint 
(\ref{v3}) an equation for $\chi \equiv \sqrt{1 - \Psi}$
\be \label{chi}
\dot \chi + H \chi 
\mp 2 \sqrt{k \lambda^{B}_{0}} \chi^2 = 0 \ .
\ee  
This is a Riccati equation and its general solution  
reads 
\be \label{ricc}
\chi (t) = e^{-\int H(t) dt} \bigl( B \mp
 \sqrt{k \lambda^B_0} \int e^{-\int H(t^\prime) dt^\prime} dt\bigr)^{-1} \ ,
\ee
where  $H(t)$ is given by solution (\ref{ah}). 
Note that $\ln \chi^{-1}$ is proportional to  the proper distance between 
the branes    
\be
d(t) = l \ln \frac{1}{\chi} \ .
\ee
Upon substituting solution (\ref{ah}) for $H$ we get
\beq \label{dtg}
d(t) &=& l \ln  \bigl(B t^{\frac{2}{3\Gamma}} \mp 
\frac{3\Gamma \sqrt{k \lambda^B_0}}{3\Gamma -2} t \bigr) \ , \quad 
\Gamma \not= \frac{2}{3} \ , \\
d(t) &=& l \ln ( \tilde{B}t \mp \sqrt{k\lambda^B_0} \ln t) \ , 
\qquad \Gamma = \frac{2}{3} 
\eeq
where $B$, $\tilde{B}$ are integration constants.
From these expressions the moments $t_c$ of the collision 
of the branes can be
determined; obviously $t_c \not= 0$. This means that the scale factor
is regular at the moment of collision and the collision
itself is nonobservable  from the viewpoint 
of A-brane observer, since it doesn't influence 
the evolution of the scale factor. 
Moment $t=0$ is singular: the distance 
between the branes tends to $-\infty$ and 
the scale factor $a(t)$ shrinks to zero (singularity).

If we choose integration constant $B$  as  
\be
B = \frac{3 \Gamma \sqrt{k \lambda_{B}^{0}}} {(3 \Gamma -2)} 
\ee
then the  expression for $\Psi = 1- \chi^2$, 
$\Gamma \not= \frac{2}{3}$ acquires a convenient form
\be \label{psi4}
\Psi = 1 - \frac{(3\Gamma - 2)^2}{9 \Gamma^2 k \lambda_{B}^{0}} 
\frac{1}{(t^{\frac{2}{3\Gamma}} \mp t)^{2}} \ .
\ee
We see that at sufficiently late times $\Psi \rightarrow 1$ and 
again the familiar general relativity appears.  
The situation is illustrated on Fig. 2. Note that the case 
$\Gamma = \frac{2}{3}$ is critical in the sense that if 
$\Gamma > \frac{2}{3}$ it is not possible to extend the time domain to 
negative values because the distance between the branes (\ref{dtg})
is not determined unless time $t$ is replaced by its absolute value 
$|t|$.

\begin{figure}[h]
%\centering
\vskip 1cm
\hskip 1.5cm 
\psfig{file=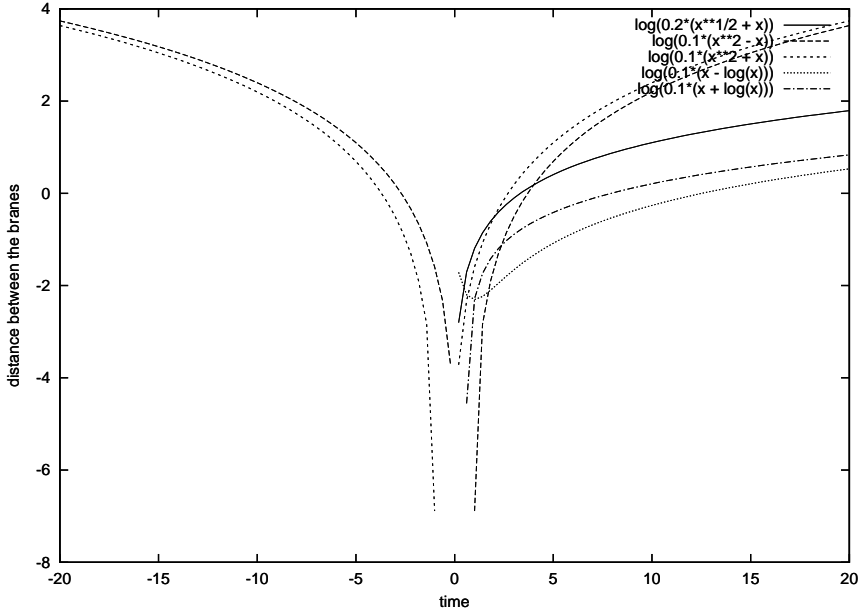}
\label{figure2}
\caption{The proper distance between the branes in the case 
of $\Gamma = \frac{4}{3}$, $\Gamma = \frac{2}{3}$
and $\Gamma = \frac{1}{3}$. Collision moments are  regular points.}
\end{figure}

As pointed out by Kanno and Soda \cite{ks1} it is possible to recover 
the 5-dimensional bulk metric from the effective 4-dimensional theory.
The 4-dimensional theory works as a hologram and this is the reason 
to call it holographic brane gravity.
Using  solutions (\ref{qi}) and (\ref{psi4}) we can write 
 the first order 
of iteration of the 5-dimensional bulk metric as follows 
($\Gamma \not= \frac{2}{3}$, $\Gamma \not= 0$) 
\be
g_{\mu\nu}(y, x^{\mu}) = (1 - \Psi)^{\frac{y}{l}} h_{\mu\nu} (x^{\mu}) =
\frac{(3\Gamma - 2)^{\frac{2y}{l}}}{ (9 \Gamma^2 k 
\lambda_{B}^{0})^{\frac{y}{l}}}
\frac{t^{\frac{4}{3\Gamma}}}
{\left(t^{\frac{2}{3\Gamma}} \mp t \right)^{\frac{2y}{l}}}  f_{ij}(x^{k})\ .
\ee
This is a ''brane-based" model of a time dependent bulk geometry 
which contains two  branes with  fixed values of $y$-coordinate \cite{k1}.
Probably it is possible to introduce an alternative ''bulk-based"
viewpoint, where the bulk remains static 
but coordinates of branes are not fixed in respect of a ''bulk-based" 
coordinate system  \cite{langlois}. Such a model
for a one brane model (a domain wall) is treated by Kraus \cite{kraus}. 
In both cases the motion of the branes will be interpreted by an observer on 
a brane as an expansion or a contraction.

\section{The influence of the dark radiation}

In general, equation (\ref{v2}) is a second order
differential equation for scale factor $a(t)$
\be
\frac{d^2 (a^2)}{dt^2} = k \rho_0 (4-3\Gamma) a^{2-3\Gamma}
\ee
and its first integral reads
\be \label{a}
\frac{d(a^2)}{\sqrt{a^{4-3\Gamma}+ \frac{C}{k\rho_0}}} =
2 \sqrt{k \rho_0} dt 
\ee
where $C$ is the constant of integration that is taken to be zero
in  solutions (\ref{dS}), (\ref{ah}).
The missing integration constant can be interpreted as the dark radiation term,
which encodes a possible influence of the 
bulk on the brane.
Eq. (\ref{a}) implies for the Hubble parameter
\be \label{h}
H^2 = \frac{k \rho_0}{a^{3 \Gamma}} + \frac{C}{a^4} 
\ee
or, taking into account  the conservation law 
 (\ref{pidevus})  
\be \label{h1}
H^2 =k \rho + \frac{C}{a^4}  \ .
\ee

In the case of vanishing dark radiation we have $C=0$
and eq. (\ref{a}) determines the same solutions as in the previous section.
In the case $\Gamma = 0 \ ,$ eq. (\ref{h}) coincides with the expression
given by Kanno et al   \cite{kss}.
  
In the case of nonvanishing dark radiation $C \not= 0 \ ,$ 
eq. (\ref{a})   can easily be integrated 
only at some special values for $\Gamma$.

1. $\Gamma = 0$ 

\be \label{a20}
a^2 = \frac{1}{2} e^{2\sqrt{k \rho_0} (t - t_0)} -
\frac{C}{2k\rho_0 e^{2\sqrt{k \rho_0} (t - t_0)}} \ .
\ee
At late times ($t \rightarrow \infty$) the 
influence of the dark radiation ($C \not= 0$) vanishes and the
solution for the scale factor acquires the familiar deSitter form. 

2. $\Gamma = \frac{4}{3}$

\be
a^2 = 2 \sqrt{k\rho_0 + C} (t-t_0) \ , \qquad H = \frac{1}{2 \sqrt{t-t_0}} \ .
\ee
We see that in the case of pure radiation matter tensor  
the dark radiation has no influence  on the Hubble parameter of the A-brane.  

3. $\Gamma = \frac{2}{3}$

\be 
a^2 = k \rho_0 (t-t_0)^2 - \frac{C}{k\rho_0} \ , \qquad
H = \frac{k \rho_0 (t - t_0)}{ a^2} \ .
\ee

4. $\Gamma = 1$

\beq
a  = \frac{1}{2} \beta^{\frac{1}{3}} + 2 C^2 \beta^{-\frac{1}{3}} + C  \ ,
\qquad
\beta  = 9 (k\rho_{0})^4 (t-t_{0})^2  - 8C^3 \nonumber 
\eeq
\be 
+ 3 \sqrt{9(k\rho_{0})^8 (t-t_{0})^4 - 16 C^3 (k\rho_{0})^4 (t-t_{0})^2} \ .
\ee

Let us now investigate constraint (\ref{v3}). 
Upon substituting  value of $k \rho$ (\ref{h1}) 
 we get an equation for $\chi \equiv \sqrt{1- \Psi}$
analogous to eq. (\ref{chi})
\be 
\dot \chi + H \chi 
\mp 2 \sqrt{k \lambda^{B}_{0} \chi^4 + \frac{C}{a^4}} = 0 \ .
\ee  

For a numerical investigation of $\Psi$ 
constraint (\ref{v3}) can be used in the following form  
\be
\frac{k \rho_0}{a^{3 \Gamma}} \frac{(1- \Psi)}{\Psi}
-H \frac{{\dot \Psi}}{\Psi} + \frac{{\dot \Psi}^2}{4 \Psi (1- \Psi)}
-k\lambda^B_0 \frac{(1-\Psi)^2}{\Psi} = \frac{C}{a^4}
\ee
where $H$ and $a$ are determined by eq. (\ref{a}).

\section{Discussion and Summary}

In this paper we considered a cosmological scenario where two branes 
are moving and colliding in a 5-dimensional bulk spacetime.
We used a low energy effective theory which is a scalar-tensor 
type theory on both branes with a specific coupling function.
The matter  is described by a barotropic 
perfect fluid on the A-brane and  
by a phenomenological time dependent ``cosmological constant''
on the B-brane.
We found some special solutions for the scale factor on the A-brane and for 
the radion which determines the proper distance between the branes.

We conclude that for all values of the barotropic index $\Gamma$, at 
late time the dynamics on the A-brane is well described by the Einstein 
general relativity with $\Psi \approx 1$.
In the case of a phenomenological cosmological constant on the A-brane 
($\Gamma =0, p = - \rho$) we have the deSitter type evolution at late time. 
This feature seems to be typical also in other braneworld scenarios discussed 
recently \cite{ds} and fits well with the experimental evidence of late time 
acceleration. Compared with the phenomenological theory (quintessence) the 
braneword model gives a more motivated theoretical ground to this result.

In the case $\Gamma \not=0$ we first have  assumed the power 
law evolution. This type of solution lacks at least one integration constant
which encodes the influence of the bulk (dark radiation on the brane).
As a result the cosmological evolution of our Universe on the A-brane
coincides with the familiar FRW model and consequently shares 
all its observational
evidences. But this also means that the solution contains no additional 
hints for approving the braneworld model. Such  hints could be found in the
explicit solutions with the dark radiation term ($C \not= 0$) presented by us
 in special cases of barotropic index  
$\Gamma = 0, \  4/3, \  2/3, \  1$. 

The dynamics of the radion is discussed in detail. We conclude that
the collision of the branes can take place at a distinct moment 
determined by matter tensors on the branes.  
The evolution of the scale factor and the radion field 
is regular at the moment of collision.
However, we have chosen a very specific coordinate system
(\cite{ks1}) and we have not discussed any other choice.
The coordinate effects must be separated from the 
physical ones and this will be a prospect for a future work.

%%%%%%%%%%%%%%%%%%%%%%%%%%%%%%%%%%%%%%%%%%%%%%%%%
\bigskip
\textbf{ACKNOWLEDGMENTS}
\smallskip

This work was supported by the Estonian Science Foundation 
under grants Nos 5026 and 4515.

%%%%%%%%%%%%%%%%%%%%%%%%%%%%%%%%%%%%%%%%%%%%%%%%%

%\newpage


\begin{thebibliography}{99}

\bibitem{rs1}
L. Randall and  R. Sundrum,  Phys. Rev. Lett. {\bf 83}, 3370 (1999).

\bibitem{rs2}
L. Randall and R. Sundrum,  Phys. Rev. Lett. {\bf 83}, 4690 (1999).

\bibitem{bdl}
P. Binetruy, C. Deffayet, and D. Langlois, Nucl. Phys. {\bf B565} (2000), 269.

\bibitem{langlois}

D. Langlois, Prog. Theor. Phys. Suppl. {\bf{148}} 181, (2003).

\bibitem{ks2}
S. Kanno and J. Soda,  Phys. Rev. {\bf D66}, 043526 (2002).

\bibitem{ks1}
S. Kanno and J. Soda,  Phys. Rev. {\bf D66}, 083506 (2002).

\bibitem{ekpyrotic}
 J.  Khoury, B.A. Ovrut, P.J.  Steinhardt,  and N. Turok,  
          Phys. Rev. {\textbf{D64}}, 123522 (2001).

\bibitem{ks}
P. Kuusk and M. Saal, Gen. Rel. Grav. {\bf 34}, 2135 (2002).

\bibitem{kss}
S. Kanno, M. Sasaki, and J. Soda,  Prog. Theor. Phys. {\bf{109}}, 357 (2003).

\bibitem{pbb}
M.   Gasperini and G. Veneziano,    Astropart. Phys. 
    \textbf{1}, 317 (1993). 

\bibitem{homepage}
    A collection of papers on the pre-big bang scenario is available at
    homepage http://www.to.infn.it/\~\ gasperin/. 

\bibitem{sa1}
A. Serena, J.M. Alimi, and A. Navarro, Class. Quant. Grav. 
{\bf 19}, 857 (2002).

\bibitem{kk}
I.M. Khalatnikov and A.Yu. Kamenshchik, Class. Quant. Grav. 
{\bf 19}, 3845 (2002).

\bibitem{k1}
N. Kaloper,  Phys. Rev. {\bf D60}, 123506 (1999).

\bibitem{kraus}
P. Kraus, J. High Energy Phys.  9912 (1999) 011.

\bibitem{ds}
S. Nojiri and S.D. Odintsov, Phys. Lett {\bf B565}, 1 (2003). 

\end{thebibliography}
\end{document}